\documentstyle[11pt,epsf,ifthen]{amsart}
\newtheorem{thm}{Theorem}[section]

\newtheorem{defn}[thm]{Definition}

\theoremstyle{remark}

\newenvironment{pr}{{\em Proof}\newline }{\epr}

\newcommand{\bR}{{\Bbb R}}

\newcommand{\cL}{{\mathcal{L}}}

\newcommand{\cO}{{\mathcal{O}}}

\newcommand{\ta}{{\tilde a}}

\newcommand{\trr}{\triangleright}
\newcommand{\epr}{\begin{flushright} $\Box$ \end{flushright}}

\begin{document}
\title{Further illustration of the use of the
Frobenius-Schwinger-Dyson equations}
\thanks{This work was financially supported by
``Samenwerkingsverband FOM/SWON Mathematische Fysica'',
under grant no. MF92.3}
\author{Olivier de Mirleau}
\address{Department of Mathematics\\ University of Amsterdam\\
Plantage Muidergracht 24\\ NL-1018 TV Amsterdam\\ The Netherlands\\
e-mail: mirleau@@wins.uva.nl}
\begin{abstract}
The Frobenius-Schwinger-Dyson equations are a rather high-brow
abstract nonsense type of equations describing $n$-point functions of
arbitrarily high composite insertions. It is not clear how to solve or even
find approximate solutions of these equations in general, but they are worth
investigating because (a certain preferred type of) renormalization of
composite insertions has been performed in advance: it just remains to find
solutions given an action and renormalization conditions.
Earlier work in this field involved only Gaussian actions or variable
transformations thereof. In this work we illustrate the use of
Frobenius-Schwinger-Dyson at a less obviously trivial level, that of
the Thirring model.

{\it Keywords}: Schwinger-Dyson equation, Frobenius Algebras, Thirring model.\\
{\it 1991 MSC}: 81 Q 40, 81 R 99, 81 T 40.
\end{abstract}

\maketitle

\tableofcontents

\section{Introduction and overview}
In a previous paper \cite{mir.compat}, we introduced the notion of
Frobenius-Schwinger-Dyson equations associated to an action. These equations
which we will remind the in next section are equations for renormalized
expectation values
$\langle \cO_1,..,\cO_n\rangle_r$ of possibly composite insertions
$\cO_i$.

Under the assumption that these expectation values are obtained as
$$\langle \cO_1,..,\cO_n\rangle_r:=
\langle R(\cO_1)..R(\cO_n)\rangle,$$
where $\langle.\rangle$ is a solution of the Schwinger-Dyson equation
associated to the action $S$, and $R$ is a regulator of composite insertions,
and if one furthermore assumes a compatibility between $R$ and $S$, one
can derive equations for the renormalized expectation values.

To get an idea of what this compatibility is, note the following:
In case $S$ is quadratic, $R$ is typically to be though of as the normal
ordering $N_S$ associated to $S$, but if $S$ is not quadratic then $R$ is
{\it not} to be thought of as normal ordering $N_0$ of
some quadratic action $S_0$,
but rather as an operation which is directly related to $S$.Now it is not
clear how to generalize normal ordering to arbitrary non-quadratic
actions $S$, so that one
is forced to think about what it means to say that $R$ is a regulator
of composite insertions directly related to $S$. The mere requirement that
infinite-dimensional calculations give finite results would reduce the
theory to triviality since we might in that case just as well take $R=0$.
In \cite{mir.compat} it was noted that normal ordering $N_S$ for
quadratic actions (or in fact a larger class for which it {\it is} clear
how to generalize normal ordering) satisfies
a compatibility condition with $S$ such that
\begin{enumerate}
\item The compatibility condition can be generalized to arbitrary action $S$.
(This is fortunate).
\item The compatibility condition does not determine $N_S$ once $S$ is
given. (Unfortunate).
\end{enumerate}

Forgetting about point two for a moment, we were then led to {\it define}
regulators as those operations $R$ that satisfy the compatibility
condition, which in turn implies identities for the corresponding
renormalized expectation values, which we called the
Frobenius-Schwinger-Dyson equations. These equations seemed so natural that
it was more or less suggested that computing functional integrals with
composite insertions and fixed action
is {\it nothing but} solving the FSD equations, although
that point was only illustrated with Gaussian integrals and variable
transformations thereof.

Finally, coming back to point two
one notes that Frobenius-Schwinger-Dyson equations
associated to an action can have multiple solutions if they are not
supplemented with extra renormalization conditions.

In this article we will push the illustration of the FSD equations a little
further by using them to derive the expression for renormalized $2n$-point
functions
$$\langle \bar\psi^{A_1}(x_1),\bar\psi^{A_2}(x_2),
\bar\psi^{A_3}(x_3), ...,
\psi^{B_1}(y_1),\psi^{B_2}(y_2),\psi^{B_3}(y_3), ...\rangle_r$$
first found by Johnson \cite{johnson}
for a two-dimensional Fermion field with the Thirring action.
To fix them, we will have to impose some renormalization conditions, which
we will choose in such a way as to remain as close as possible to
Johnson's expressions. What Johnson's method amounts to
in our setting is to guess the form of renormalization conditions
that might allow for solutions of the FSD equations, and subsequently
fixing free parameters in this guess just by the very
imposition of these FSD equations.

\section{Reminder on the Frobenius-Schwinger-Dyson equation}
\setcounter{subsection}{1}
We recall the following from \cite{mir.compat}:
\begin{defn}
Let $A$ be a symmetric associative algebra with unit and typical
elements $f,g$, let $L:=Der(A)$ with typical elements $X,Y$ and
corresponding operations
$fg$, $fX$, $Xf$ and $[X,Y]$.
By a renormalized structure on $A$ we mean the datum of $S\in A$,
together with extra multiplications
$f._r g$, $f._r X$, $X._r f$ and $[X,_r Y]$, also satisfying associativity,
Jacobi identity etc., such that
\begin{enumerate}
\item The $._r$-operations induce $L=Der(A,._r)$, i.e.~the derivations of the
multiplication $(f,g)\mapsto f._rg$ are exactly given by the operations
$f\mapsto X._r f$ as $X\in L$.
\item Both structures have the same unit: $1f=f=1._r f$.
\item The two algebraic structures and the action $S$ are compatible in the
sense that:
$$X._r (YS)-Y._r (XS)=[X,_r Y] S,$$
$$(f._r X)S=f._r (XS)-X._r f.$$
\end{enumerate}
The algebra $A$ together with the renormalized structure will be
called a renormalized volume manifold.
For symmetric linear maps $\langle,,,\rangle_r: A^{\otimes n}
\rightarrow \bR$ we will be interested in the following properties:
\begin{enumerate}
\item $\langle .,. \rangle_r:A^{\otimes 2}\rightarrow \bR$ is a positive
non degenerate form. (Positivity).
\item $\langle \cO_1,\cO_2,...,\cO_n\rangle_r=
\langle \cO_1._r \cO_2,..,\cO_n\rangle_r,$ (Frobenius).
\footnote{A Frobenius algebra is a symmetric associative algebra with metric
such that $(a,bc)=(ab,c)$.}
\item$\langle XS,\cO_1,...,\cO_n\rangle_r=
\sum_{i=1}^n  \langle \cO_1,.., X._r \cO_i,..,\cO_n\rangle_r.$
(Schwinger-Dyson).
\end{enumerate}
\end{defn}
By substituting $S/\hbar$ for $S$ and taking the limit $\hbar\rightarrow 0$,
we see that at $\hbar=0$, we may take both algebraic structures to be
equal. In what follows, we will take the usual multiplication to have
priority over the renormalized one, i.e.~$f._r gh$ means $f._r (gh)$.

In this article we will concern ourselves with the case where $A$ is the
algebra of words in odd symbols $\psi^A(x),\bar \psi^A(x)$ and their
derivatives, and $S$ is given by the Thirring action.
\section{Reminder on chiral currents}
\setcounter{subsection}{1}
\subsubsection{Chiral currents.}
By a chiral current we shall understand a current $J$ which is divergenceless
and rotationfree if the Euler-Lagrange equations are
satisfied\footnote{In two dimensions this reduces to the more common meaning
of a current $J=*J_\mu dx^\mu=fdz+g d\bar z$ being chiral
iff $f$ is holomorphic and
$g$ antiholomorphic.}.
More precisely, currents such that there are functional vector fields $Y$ and
$Y_{\mu\nu}$ such that

$$\partial_\mu J^\mu = Y S,$$
$$\partial_\mu J_\nu - \partial_\nu J_\mu = Y_{\mu\nu} S.$$

In the context of renormalized volume manifolds, one may now define
$[J^\mu\trr_r . ]:A\rightarrow A$
by the following properties:

$$\partial_\mu [J^\mu\trr_r \cO] := Y ._r \cO,$$
$$\partial_\mu[J_\nu\trr_r \cO]
- \partial_\nu[J_\mu \trr_r \cO] := Y_{\mu\nu} ._r\cO.$$
$$[J^\mu(\infty),\trr_r \cO]:= 0.$$

Note that this differential equation for $[J^\mu\trr_r . ]$
can be solved rather explicitly in terms of an integral expression
of the right hand side.

We now see that if $\langle J^\mu(\infty), \cO\rangle_r=0$ then the
chiral currents satisfy the identity
$$\langle J^\mu,\cO \rangle_r =
\langle [J^\mu \trr_r \cO ] \rangle_r .$$

Such an identity may in turn be used to produce differential equations
for $n$-point functions in case certain identities are known for the
$._r$ operations. Such identities may be known in the following cases:
\begin{enumerate}
\item Everything was already solved by some other method, say because $S$
was Gaussian.
\item One guesses properties of $._r$.
\end{enumerate}

\subsubsection{Example 1} As a first example consider the action
$S(\phi)=\int \partial_\mu \phi \partial^\mu \phi d^D x/2$ for scalar fields
$\phi:\bR^D\rightarrow \bR$, and renormalization condition
$$\cO_1._r
\frac{\delta}{\delta \phi(x)}
\cO_2
=
\cO_1
\frac{\delta}{\delta \phi(x)}._r
\cO_2.$$
This renormalization condition was discussed in more detail
in \cite{mir.compat}: The FSD equations for non degenerate Gaussian integrals
subjected to this condition allow for a unique solution
usually referred to as
{\it the} solution for this action. The solution of the $._r$
multiplication in terms of the usual one can for this renormalization
condition and action $S$ be inductively expressed as follows
(see appendix, proof \ref{opeproof}):
$$(\frac{\delta S}{\delta \phi(x)}\cO_1)._r \cO_2
=\frac{\delta S}{\delta \phi(x)}(\cO_1._r \cO_2)
+
\cO_1._r \frac{\delta }{\delta \phi(x)} \cO_2.$$
Now the action above also happens to have a chiral current $j_\mu:=
\partial_\mu\phi$, by taking $Y^x:=-\delta/\delta \phi(x)$, and
$Y_{\mu\nu}:=0$. The expression $V_p(x):=\exp(p\phi(x))$ then
has the  property that its derivative can be algebraically expressed
in terms of $j_\mu$ and $V_p$ itself:
$$\partial_\mu V_p = p j_\mu V_p.$$
The properties of $._r$ that are needed to
derive a differential equation for $n$-point functions of the $V_p$'s are
\begin{enumerate}
\item $Y^x._r V_p(y)=Y^x V_p(y)$,
\item $j_\mu(x)V_p(x)=
\lim_{y\rightarrow x} j_\mu(y)._r V_p(x)-[j_\mu(y)\trr_r V_p(x)].$
\end{enumerate}
Indeed, the first equation above follows directly from the main renormalization
condition; For the second equation we use
$j_\mu(x) V_p(x)=\overline\lim_{\epsilon\rightarrow 0} j_\mu(y)V_p(x)$,
and use the Gaussian recursion relation for the $._r$ product.

The way these two equations imply a differential equation for $n$-point
functions is as follows:
$$\partial_{x^\mu} \langle V_p(x),\cO\rangle_r =
p\langle j_\mu(x) V_p(x),\cO\rangle_r
=
p\lim_{y\rightarrow x}
\langle (j_\mu(y)._r V_p(x)-[j_\mu(y)\trr_r V_p(x)]),\cO\rangle_r$$
$$=
p\lim_{y\rightarrow x}
\langle j_\mu(y), V_p(x)._r \cO\rangle_r
-\langle [j_\mu(y)\trr_r V_p(x)],\cO\rangle_r
=\lim_{y\rightarrow x} 
\langle V_p(x),[j_\mu(y)\trr_r \cO]\rangle_r$$
$$=\langle V_p(x),[j_\mu(x)\trr_r \cO]\rangle_r$$
\subsubsection{Example 2:} Next consider a slightly more complicated case,
that of Fermionic spinors $\psi$ and $\bar \psi$, with Dirac action
$S=\int \bar \psi(\gamma^\mu \partial_\mu -im)\psi d^D x$,
together with
the Fermionic analogue of the renormalization condition of the previous
example. This case is just as solvable as the previous one since it
is also Gaussian. The current $j_\mu:=\bar \psi \gamma_\mu \psi$ is chiral
in case we have $m=0$ and $D=2$: Indeed defining the functional vector fields
$$Y^x:=
(\bar \psi^A(x) \frac{\delta}{\delta \bar \psi^A(x)}
-\psi^A(x) \frac{\delta}{\delta \psi^A(x)}),$$
$$Y_{\mu\nu}:=\frac{1}{2}
(\psi^A {[\gamma_\mu,\gamma_\nu]_A}^B
\frac{\delta}{\delta \psi^B}
-\bar\psi^A {[\gamma_\mu,\gamma_\nu]_A}^B \frac{\delta}{\delta \bar\psi^B}),$$
we have
$$\partial_\mu j^\mu = YS,$$
$$\partial_\mu j_\nu - \partial_\nu j_\mu - Y_{\mu\nu} S
= \frac{1}{2}\{\partial_\sigma \bar \psi \Gamma_{\mu\sigma\nu} \psi
-\bar \psi \Gamma_{\mu\sigma\nu} \partial_\sigma \psi
+2im \bar \psi [\gamma_\mu,\gamma_\nu]\psi \},$$
\[ where\;\;\Gamma_{\mu\sigma\nu}:=\gamma_\mu \gamma_\sigma \gamma_\nu
-\gamma_\nu \gamma_\sigma \gamma_\mu,\]
which is zero in $D=2$.
Assuming that $m=0$ and $D=2$, one may again derive properties of $._r$
that imply a differential equation, for $n$-point functions $\psi$ and
$\bar \psi$'s this time: Setting $\varepsilon_{\mu\nu}\tilde Y:=Y_{\mu\nu}$,
where $dx^\mu \wedge dx^\nu=:\varepsilon^{\mu\nu}dx^1 \wedge dx^2$,
and using the notion
$$\overline\lim_{\epsilon\rightarrow (0,0)\in \bR^2} \;f(\epsilon)
:=\lim_{\epsilon\rightarrow 0} \;\frac{1}{4}
\sum_{i=0}^3 f(Rot_{i\times 90^o}(\epsilon)),$$
so that in particular
$\overline\lim_{\epsilon\rightarrow 0} \;\epsilon_\alpha f(|\epsilon|^2) =0$,
and
$\overline\lim_{\epsilon\rightarrow 0}
\;\frac{\epsilon^\alpha \epsilon^\beta}{|\epsilon|^2}=
\frac{g^{\alpha\beta}}{2}$,
we have:
\begin{enumerate}
\item $Y._r \psi = Y \psi$ ; $Y._r \bar \psi = Y \bar \psi$ ;
$\tilde Y._r \psi = \tilde Y \psi$ ;
$\tilde Y._r \bar \psi = \tilde Y \bar \psi $
\item $\overline\lim_{\epsilon \rightarrow 0}^{~}
\{ j_\mu(x+\epsilon)._r \psi^A(x) - [ j_\mu(x+\epsilon)\trr_r \psi^A(x)]
-j_\mu(x+\epsilon)\psi^A(x)\}$
$$=\frac{-1}{4\pi} {(\gamma^\nu\gamma_\mu)^A}_B \partial_\nu \psi^B(x).$$
(See appendix, proof \ref{diracproof}).
\item $\frac{\delta}{\delta \psi^A(x)}._r \psi^B(y) = \delta_A^B \delta(x-y)$.
\item $j^\mu(x)=\overline\lim_{\epsilon\rightarrow 0}\;
\bar \psi^A(x)._r (\gamma_\mu)_{AB} \psi^B(x+\epsilon)$.
\end{enumerate}

That these properties by themselves imply a differential equation for the
$2n$-point functions will be shown in more generality in the next section,
from which the present case is recovered by setting $\lambda:=0$.
\section{The Thirring model with Johnson's renormalization conditions}
\setcounter{subsection}{1}
In the associative algebra on odd symbols
$\psi^A(x)$ and $\bar \psi^A(x)$ where $x\in \bR^2$ and
${}^A$ is an index in a $2D$ representation space for the Clifford
algebra of $\bR^2$ with $\gamma_{AB}^\mu=\gamma_{BA}^\mu$,
Thirring's Lagrangian is defined
as $\cL(\bar \psi,\psi):=
\bar \psi \gamma^\mu \partial_\mu \psi +\frac{\lambda}{2} j^\mu j_\mu$,
where $j^\mu:=\bar \psi \gamma^\mu \psi$.
Consequently, we have
\begin{enumerate}
\item
$\frac{\delta \cL}{ \delta \bar \psi^A} =
(\gamma^\mu)_{AB} \partial_\mu \psi^B
+\lambda (\gamma^\mu)_{AB}j_\mu\psi^B$.
\item
$\frac{\delta \cL}{ \delta \psi^A} =
\partial_\mu \bar \psi^B (\gamma^\mu)_{BA} - 
\lambda (\gamma^\mu)_{AB}j_\mu\bar \psi^B$.
\item $\partial_\mu j^\mu = YS$
\item $\partial_\mu j_\nu- \partial_\nu j_\mu = Y_{\mu\nu}S$.
\end{enumerate}
I.e.~although the Lagrangian is not quadratic any more, the current $j_\mu$
remains chiral even when $\lambda\neq 0$,
which basically makes a solution of the Thirring model possible.

Next, we need renormalization conditions to fix the renormalized
$2n$-point functions. Now it is not evident what kind of conditions can
be imposed such that solutions of the Frobenius-Schwinger-Dyson equations
satisfying them actually exist. This general problem will remain unaddressed
in this paper: We will restrict ourselves to deriving a number of
Johnson's results \cite{johnson}
by using the Frobenius-Schwinger-Dyson equations:
In our language, Johnson made a smart guess about the form of
a number of renormalization
conditions, by introducing some undetermined
parameters $a,\ta$ in the conditions as they hold in the Gaussian $\lambda=0$
case:

\begin{enumerate}
\item $Y._r \psi = a Y \psi$ ; $Y._r \bar \psi = aY \bar \psi$ ;
$\tilde Y._r \psi = \ta\tilde Y \psi$ ;
$\tilde Y._r \bar \psi = \ta\tilde Y \bar \psi$,
see \cite[formulae 14,15]{johnson}.
\item
\label{bcond}
$\overline\lim_{\epsilon \rightarrow 0}
\{ j_\mu(x+\epsilon)._r \psi^A(x) - [ j_\mu(x+\epsilon)\trr_r \psi^A(x)]
-j_\mu(x+\epsilon)\psi^A(x)\}$
$$=\frac{-b}{4\pi} {(\gamma^\nu\gamma_\mu)^A}_B \partial_\nu \psi^B(x).$$
\item $x\neq y\;
\Rightarrow\;
\frac{\delta}{\delta \psi^A(x)}._r \psi^B(y)=0$\footnote{We will not address
the question whether this condition can be consistently extended to
be a distribution for all values of $x$ and $y$ including $x=y$.
The price we pay is that solutions of the differential equations that follow
need not be unique.}.
\item $j^\mu(x)=\overline\lim_{\epsilon\rightarrow 0}\;
U_{\lambda,a,\ta}(\epsilon^2)
\bar \psi^A(x)._r (\gamma_\mu)_{AB} \psi^B(x+\epsilon)$,
see \cite[formula 29]{johnson} and \cite[formula 4.71]{wightman},
\label{currenoco}
\end{enumerate}
where
$U_{\lambda,a,\ta}$ is defined by
$\langle \bar \psi^A(\epsilon),\psi^B(0)\rangle_r^{\lambda,a,\ta}
=U_{\lambda,a,\ta}^{-1}(\epsilon^2)
\langle \bar \psi^A(\epsilon),\psi^B(0)\rangle_r^{\lambda=0,a=1,\ta=1}$, which
as we will see exists.
We see that
the Gaussian properties with canonical renormalization condition are
recovered by setting $a=\ta=b=1$.
The parameter $b$ does not occur in Johnson's work, but we will need
condition \ref{bcond}
to make everything work. It will not enter the final answer because
the $b$-term will be multiplied by $\gamma^\mu$, and since in two
dimensions $\gamma^\mu \gamma_\nu\gamma_\mu=0$.

Repeating the procedure that we have seen before to produce
differential equations,
we see that the defining differential equation for $[j_\mu\trr_r .]$
acting  on $\psi,\bar \psi$
is solved by
$$[j^\mu(x)\trr_r \psi^A(y)] =
(- a f^\mu (x-y) \delta_B^A + \ta \varepsilon^{\mu\nu} f_\nu(x-y)
\bar \gamma^A{}_B) \psi^B(y),$$
$$[j^\mu(x)\trr_r \bar\psi^A(y)] =
(a f^\mu (x-y) \delta_B^A - \ta \varepsilon^{\mu\nu} f_\nu(x-y)
\bar \gamma^A{}_B) \bar\psi^B(y),$$
where $\bar \gamma:=\gamma^1 \gamma^2$,
$f(x) :=(\ln|x|)/2\pi$, and
$f_\mu(x):=\partial_\mu f(x)$. For example:
$$\partial_\mu[j^\mu(x)\trr_r \psi^A(y)] =
- a \partial_\mu f^\mu (x-y) \delta_B^A \psi^B(y)
=-a \delta(x-y) \delta_B^A \psi^B(y) = a Y^x \psi^A(y)=Y^x ._r \psi^A(y).$$

The differential equation that now holds is the following if $\cO$ is
a repeated $._r$-product of $\psi(y_i),\bar\psi(y_i)$'s where $x\neq y_i$:
$$\langle (\gamma^{\mu})^A{}_B \partial_\mu\psi^B(x), \cO\rangle_r
=
- \lambda (\gamma^{\mu})^A{}_B
\langle\psi^B(x), [j_\mu(x)\trr_r\cO]\rangle_r.$$
\begin{pr}
$$LHS=\langle \frac{\delta \cL}{\delta \bar \psi^A(x)}
-\lambda j_\mu(x)(\gamma^\mu \psi)^A(x),\cO\rangle
=-\lambda (\gamma^\mu)^A{}_B \langle j_\mu(x) \psi^B(x),\cO\rangle$$
$$=- \lambda (\gamma^\mu)^A{}_B\overline\lim_{\epsilon\rightarrow 0}
\langle j_\mu (x+\epsilon)._r \psi^B(x) - [ j_\mu(x+\epsilon)\trr_r \psi^B(x)]
+\frac{b}{4\pi} (\gamma^\nu\gamma_\mu)^B{}_C \partial_\nu
\psi^C(x),\cO\rangle$$
$$=-\lambda (\gamma^\mu)^A{}_B\langle \psi^B(x),[j_\mu(x)\trr_r \cO]\rangle
-\frac{\lambda b}{4\pi}(\gamma^\mu \gamma^\nu \gamma_\mu)^A{}_B
\langle \partial_\nu \psi^C,\cO\rangle.$$
\end{pr}

Up to now, renormalization condition \ref{currenoco} has not been used.
We will prove in the next section that it implies the relations first
found by Johnson
$$a=\frac{1}{1-\lambda/2\pi},\;\;\ta=\frac{1}{1+\lambda/2\pi}.$$

\section{Free field realization, restriction of the parameters $a,\ta$.}
\setcounter{subsection}{1}
\label{freesec}
The differential equation satisfied by the expectation values
of the Thirring model can be solved by expectation values of special
integrands of a Gaussian weight: Indeed, define the following Gaussian
contractions \cite{mir.contr}
for two Bosonic fields $\phi_1$ and $\phi_2$
and Dirac fields $\chi$ and $\bar \chi$:
$$ [\phi_i(x) \trr \phi_j(y)]:=-\delta_{ij} f(x-y),$$
$$[\chi\trr \chi]:=[\bar \chi \trr \bar \chi]:=0,$$
$$[\chi^A(x)\trr \bar \chi^B(y)]:=(\gamma^\mu)^{AB} f_\mu(x-y)
=:[\bar \chi^A(x)\trr \chi^B(y)].$$
Following \cite[formula IV.5]{klaiber}, let us define
$$\Psi^A(x):= (e^{-\lambda k_1 \phi_1(x) - \lambda k_2 \phi_2(x)
\bar \gamma})^A{}_B
\chi^B(x),$$
$$\bar\Psi^A(x):= (e^{+\lambda k_1 \phi_1(x)  +\lambda k_2 \phi_2(x)
\bar \gamma})^A{}_B
\bar\chi^B(x),$$
for yet to be fixed numbers $k_1, k_2$. Making the Dirac operator
$\gamma^\mu\partial_\mu$ act on these expressions will bring down the matrix
$\bar \gamma$, and using that $\gamma^\mu \bar \gamma =
\varepsilon^{\mu\nu}\gamma_\nu$, we become interested in
$$j_\pm^\mu:=k_1 \partial^\mu \phi_1 \pm k_2 \varepsilon^{\mu\nu}
\partial_\nu \phi_2$$
By Gaussian Wick calculus we then have:
$$[j_-^\mu(x) \trr \Psi^A(y)]
=\{\lambda k_1^2 f^\mu(x-y) \delta_B^A - \lambda k_2^2
\varepsilon^{\mu\nu} f_\nu(x-y)
\bar \gamma^A{}_B\} \Psi^B(y),$$
$$[j_-^\mu(x) \trr \bar \Psi^A(y)]
=\{-\lambda k_1^2 f^\mu(x-y) \delta_B^A + \lambda k_2^2
\varepsilon^{\mu\nu} f_\nu(x-y)
\bar \gamma^A{}_B\} \bar\Psi^B(y),$$
$$\frac{\delta\cO}{\delta\bar\psi(x)}=0\Rightarrow
(\gamma_\mu)^A{}_B \langle \partial_\mu \Psi^B(x),\cO\rangle_r
=-\lambda
(\gamma_\mu)^A{}_B \langle \Psi^B(x),[j_-^\mu(x)\trr \cO]\rangle_r.$$
The last differential equation is exactly the one derived using
Johnson's renormalization conditions, provided the action $[j_-^\mu\trr .]$
equals the action of $j^\mu$, which happens
if we have $-\lambda k_1^2 = a$ and
$-\lambda k_2^2 = \ta$. Consequently the $n$-point functions of the
$\Psi$'s above with these values of $k_i$ satisfy the differential
equation for the $\psi$'s of the Thirring $n$-point functions.
In particular the two-point function can be computed, leading to
$$\langle \bar \psi^A(\epsilon),\psi^B(0)\rangle_r^{\lambda,a,\ta}
=e^{\lambda (a-\ta) \langle \phi(0)\phi(\epsilon)\rangle}
\langle \bar \psi^A(\epsilon),\psi^B(0)\rangle_r^{\lambda=0,a=1,\ta=1},$$
so that
$U_{\lambda,a,\ta}^{-1}(\epsilon^2)
=e^{\lambda (a-\ta) \langle \phi(0)\phi(\epsilon)\rangle}$.
Furthermore, in view of renormalization condition \ref{currenoco}
the Thirring expectation values of $j_\mu(x)$ are reproduced by the following
expression in terms of the Gaussian fields:
$$J^\mu(x):=\overline\lim_{\epsilon\rightarrow 0}\;
U_{\lambda,a,\ta}(\epsilon^2)
\bar \Psi^A(x)._r (\gamma_\mu)_{AB} \Psi^B(x+\epsilon).$$
But using Gaussian Wick calculus, one may prove (see appendix, proof
\ref{currentproof}) that
$$J^\mu = \bar \Psi^A(\gamma^\mu)_{AB} \Psi^B
+\frac{\lambda}{2\pi} j_+^\mu
= \bar \chi^A(\gamma^\mu)_{AB} \chi^B
+\frac{\lambda}{2\pi} j_+^\mu
=:j^\mu_\chi+\frac{\lambda}{2\pi} j_+^\mu.$$
Since $\langle j^\mu, \cO\rangle_r = \langle [j^\mu\trr_r\cO]\rangle_r$,
and comparing $[j^\mu\trr_r\psi^A]$ with
$[j^\mu_\chi+\frac{\lambda}{2\pi} j_+^\mu\trr \Psi^A]$, we see that the
following two expressions must be equal:
\begin{enumerate}
\item $(- a f^\mu (x-y) \delta_B^A + \ta \varepsilon^{\mu\nu} f_\nu(x-y)
\bar \gamma^A{}_B) \psi^B(y),$
\item $(- (1+\frac{\lambda a}{2\pi}) f^\mu (x-y) \delta_B^A
+ (1-\frac{\lambda\ta}{2\pi}) \varepsilon^{\mu\nu} f_\nu(x-y)
\bar \gamma^A{}_B) \psi^B(y),$
\end{enumerate}
giving that
$a=\frac{1}{1-\lambda/2\pi}\;\;\ta=\frac{1}{1+\lambda/2\pi}$ as promised.

\section{Partial vs. Total renormalization conditions.}
\setcounter{subsection}{1}
By partial renormalization conditions we shall understand
renormalization conditions which together with the FSD equation do not
completely determine the $._r$ operations, as opposed to total
renormalization conditions which will be those which do fix the $._r$'s:
For example $f\partial_i ._r g =f ._r \partial_i g$
is a total renormalization
condition for Gaussian actions.

Johnson's renormalization conditions for the Thirring model seem to be
only partial, since when restricted to the Gaussian $\lambda=0$ case,
Johnson's conditions are obviously weaker than
$f\partial_i ._r g =f ._r \partial_i g$.

One may ask if there is a natural way to strengthen Johnson's renormalization
conditions so that they also determine the operator product or
expectation values of higher order insertions. A simple try at guessing
such an extension is to first extend the free field realization
$F: \psi\mapsto \exp(-\lambda k_1\phi_1-\lambda k_2\phi_2 \bar\gamma)
\psi_0$ by $F(ab):=F(a)F(b)$,
and then trying to pull-back the Gaussian $._r$ operations by $F$.
This however fails because $F(a)._r F(b)$ is not in the image of $F$
for all $a$ and $b$.
Another try could be to use the bosonization procedure as a free
field realization. The same problems as above might in that case
occur although we didn't check that.

\section{Conclusion}
What we have seen is that the Frobenius-Schwinger-Dyson equations can
reproduce known expressions for renormalized expectation values, the
advantage being that there was no need of explicitly regularizing and
renormalizing the fields. It remains to be proved that there actually
exists a solution of the FSD equations for the Thirring model.

It would seem that the same method will apply for example to the
Wess-Zumino-Witten model, since the derivation \cite{knizhnik}
of differential equations for $n$-point functions in
that case is not different from Johnson's derivation:
Arbitrary multiplicative parameters $\kappa=\bar\kappa$ have been introduced
\cite[formula 3.1]{knizhnik}, and consistency with renormalization
conditions on higher composite operators fixes the value of $\kappa$ in
terms of the integer coupling constant $k$ of the model
\cite[formula 3.11]{knizhnik}.

>From a more fundamental point of view it remains an open question,
even in finite dimensions, whether
there is an algorithm which given an action and renormalization conditions
will produce approximations of solutions of the FSD equations
up to arbitrarily high precision.
Unlike approximation schemes for positive solutions of the usual
Schwinger-Dyson equation,
which are just the approximation schemes for integration,
an approximation scheme for solutions of the FSD equations seems difficult
to construct given that
the number of solutions will in general depend on the
renormalization conditions.
\section{Acknowledgements}
Thanks are due to R.~H.~Dijkgraaf, H.~G.~B.~Pijls and C.~D.~D.~Neumann for
fruitful discussions.

\appendix
\section{Proofs}
\setcounter{subsection}{1}
\begin{enumerate}
\item
\label{opeproof}
The renormalization condition $f._r\partial_i g= f\partial_i ._r g$
implies $(f._r g)\partial_i = f._r (g\partial_i)$
and $((\partial_i S)\cO_1)._r \cO_2 = (\partial_i S) (\cO_1._r \cO_2)
+ \cO_1 ._r \partial_i \cO_2$:

\begin{pr} To prove the first formula, we have to show that both sides of the
equation have the same $._r$ action:
$$[(f._r g)\partial]._r h = (f._r g)._r \partial h
=f._r (g._r \partial h)
=f._r (g\partial._r h)
=(f._r g\partial)._r h.$$
The second formula now follows from the first as follows:
In the formula $f ._r XS - X ._r f=(f ._r X)S$ that holds by definition
of renormalized volume manifolds, take $f:=\cO_2$, and $X:=\cO_1 \partial_i$.
This gives
$$\cO_2 ._r (\cO_1 \partial_i S) - \cO_1 ._r \partial_i \cO_2
=
\cO_2 ._r (\cO_1 \partial_i S) - \cO_1 \partial_i._r \cO_2$$
$$=f._r XS - X ._r f = (f ._r X)S = (\cO_2 ._r \cO_1\partial_i)S$$
By the first formula, the last term equals $(\cO_2._r\cO_1)\partial_i S$.
\end{pr}
\item
\label{diracproof}
For Gaussian Dirac fields we have:
$$\overline\lim_{\epsilon \rightarrow 0}
\{ j_\mu(x+\epsilon)._r \psi^A(x) - [ j_\mu(x+\epsilon)\trr_r \psi^A(x)]
-j_\mu(x+\epsilon)\psi^A(x)\}$$
$$=\frac{-1}{4\pi} {(\gamma^\nu\gamma_\mu)^A}_B \partial_\nu \psi^B(x).$$

\begin{pr}
Indeed, the left hand side is the limit of
$$[\psi^A(x)\trr j_\mu(x+\epsilon)]
- [ j_\mu(x+\epsilon)\trr_r \psi^A(x)]$$
$$=[\psi^A(x)\trr \bar \psi^B(x+\epsilon)](\gamma_\mu)_{BC} \psi^C(x+\epsilon)
+f_\nu(\epsilon)(\gamma^\nu \gamma_\mu)_{AC} \psi^C(x)$$
$$=-(\gamma^\rho \gamma_\mu)_{AC} f_\rho(\epsilon) 
\{\psi^C(x+\epsilon)-\psi^C(x)\}
=-(\gamma^\rho \gamma_\mu)_{AC}\frac{\epsilon_\rho\epsilon_\nu}{2\pi\epsilon^2}
\partial^\nu \psi^C(x) + O(\epsilon),$$
which in the limit approaches right hand side.
\end{pr}
\item
\label{currentproof}
The different currents introduced in section \ref{freesec} satisfy the
following identity:
$$J^\mu = \bar \Psi^A(\gamma^\mu)_{AB} \Psi^B
+\frac{\lambda}{2\pi} j_+^\mu
= \bar \chi^A(\gamma^\mu)_{AB} \chi^B
+\frac{\lambda}{2\pi} j_+^\mu
=:j^\mu_\chi+\frac{\lambda}{2\pi} j_+^\mu.$$
\begin{pr}
First by expanding roughly speaking
$\Psi=e^{-(...)} \chi$ and $\bar\Psi = e^{+(...)} \bar \chi$, so that
$\bar\Psi._r\Psi= (e^+ \bar \chi) ._r (e^- \chi) = \{e^+\}._r \{e^-\}
(\bar \chi \chi + \langle\bar \chi\chi\rangle)$, and using the
matrix version of $e^{px}._r e^{qx} = e^{pq} e^{(p+q)x}$, we have:
$$e^{-\lambda a \langle \phi(x)\phi(y)\rangle
 - \lambda \ta\langle \phi(x)\phi(y)\rangle \bar\gamma_x\bar\gamma_y}
(\bar\Psi(x)._r \Psi(y))$$
$$
=
\bar\Psi(x) \Psi(y)
+
e^{\sqrt{-\lambda a}(\phi_1(x)-\phi_1(y)) + \sqrt{-\lambda \ta}
(\phi_2(x) \bar \gamma_x - \phi_2(y) \bar \gamma_y)}
\langle \bar\chi(x)\chi(y)\rangle,$$
where $\bar \gamma_x$ acts on $\bar\chi(x)$, i.e.
$(\bar\gamma_x \bar\chi(x) \chi(y))^{AB}=(\bar\gamma^A{}_C
\bar\chi^C(x) \chi^B(y))$.
>From this formula one deduces the identity among currents by taking the limit
$y\rightarrow x$:
$$j_\chi^\mu(x) =\bar\Psi^A(x)\gamma_{AB}^\mu\Psi^B(x)
=\overline\lim_{\epsilon \rightarrow 0}\;
\bar\Psi^A(x) \gamma_{AB}^\mu \Psi^B(x+\epsilon=:y)= I-II,$$
where with $B_\epsilon:=\langle \phi(0)\phi(\epsilon)\rangle$:
$$I
=\overline\lim_{\epsilon \rightarrow 0}
(e^{-\lambda a B_\epsilon -
\lambda \ta B_\epsilon\bar\gamma_x\bar\gamma_y})^A{}_C{}^B{}_D
\bar\Psi^C(x)._r \Psi^D(y) \gamma_{AB}^\mu$$
$$=\overline\lim_{\epsilon \rightarrow 0}
e^{-\lambda(a-\ta)B_\epsilon}
\bar\Psi^C(x)._r  \gamma_{CD}^\mu\Psi^D(y)=J^\mu(x)$$
$$II
=\overline\lim_{\epsilon \rightarrow 0}\;
[e^{\sqrt{-\lambda a}(\phi_1(x)-\phi_1(y)) + \sqrt{-\lambda \ta}
(\phi_2(x) \bar \gamma_x - \phi_2(y) \bar \gamma_y)}]^A{}_C{}^B{}_D
(\gamma^\rho)^{CD} f_\rho(x-y) (\gamma^\mu)_{AB}$$
$$=\overline\lim_{\epsilon \rightarrow 0}\;
e^{\sqrt{-\lambda a}(\phi_1(x)-\phi_1(y))}
f_\rho(x-y)
Tr(e^{\sqrt{-\lambda\ta}(\phi_2(x)-\phi_2(y))
\bar\gamma}\gamma^\rho\gamma^\mu)$$
$$=\overline\lim_{\epsilon \rightarrow 0}\;
e^{\sqrt{-\lambda a}(\phi_1(x)-\phi_1(y))}f_\rho(x-y)
$$
$$\times\{2g^{\rho\mu}\cos[\sqrt{-\lambda a}(\phi_1(x)-\phi_1(y))]
-2\varepsilon^{\rho\mu}\sin[\sqrt{-\lambda a}(\phi_1(x)-\phi_1(y))]\}$$
$$=\overline\lim_{\epsilon \rightarrow 0}\;
[1-\sqrt{-\lambda a}\epsilon^\kappa\partial_\kappa \phi_1(x) + O(\epsilon^2)]
\frac{-\epsilon_\rho}{2\pi \epsilon^2}2
[g^{\rho\mu} ( 1 + O(\epsilon^2))+\varepsilon^{\rho\mu}
\sqrt{-\lambda a}\epsilon^\sigma \partial_\sigma \phi_2(x) + O(\epsilon^2)]$$
$$=\frac{1}{2\pi}(\sqrt{-\lambda a} \partial_\rho \phi_1(x) g^{\rho\mu}
-\varepsilon^{\rho\mu} \sqrt{-\lambda \ta} \partial_\rho \phi_2(x))
=\frac{\lambda}{2\pi} j_+^\mu(x)$$
\end{pr}
\end{enumerate}

\end{document}